# Generation of Sound Bullets with a Nonlinear Acoustic Lens

Alessandro Spadoni,[1] and Chiara Daraio[1,2*]

[1]Graduate Aerospace Laboratories (GALCIT)
[2]Division of Applied Physics
California Institute of Technology,
Pasadena, CA, 91125, USA

*To whom correspondence should be addressed; E-mail: daraio@caltech.edu.

**Acoustic lenses are employed in a variety of applications, from biomedical imaging and surgery, to defense systems, but their performance is limited by their linear operational envelope and complexity. Here we show a dramatic focusing effect and the generation of large amplitude, compact acoustic pulses (sound bullets) in solid and fluid media, enabled by a tunable, highly nonlinear acoustic lens. The lens consists of ordered arrays of granular chains. The amplitude, size and location of the sound bullets can be controlled by varying static pre-compression on the chains. We support our findings with theory, numerical simulations, and corroborate the results experimentally with photoelasticity measurements. Our nonlinear lens makes possible a qualitatively new way of generating high-energy acoustic pulses, enabling, for example, surgical control of acoustic energy.**

The ability to accurately measure acoustic waves and control their generation and propaga-tion through a medium is of considerable interest in medical diagnosis, nondestructive inspection of materials, and remote sensing (*1*), among a variety of other applications. Acoustic signals, for example, enable ultrasonic transducers to image the interior of the human body (*2, 3*). They can also be used as a non-intrusive scalpel for surgery via beamforming or wave focusing (*4*). Similar techniques are employed to detect the presence of defects in solids without damaging the tested sample (*5*).

In these applications, acoustic waves, usually generated with electromechanical transducers, are focused at a desired location by means of geometric focusing (*5*), time reversal methods (*6*), or beamforming via phase lags (*4, 7, 8*). In the first case, the transducers' geometry is exploited to direct signals, often resulting in cumbersome and application-specific devices. In the second and third case, actuators are employed to generate a localized signal at a given location via appropriate phase delays. Each technique is limited by its reliance on electromechanical transducers which cannot produce compact, non-oscillatory or high-amplitude signals, and by the need for external electronic equipment. These requirements reduce their spatial accuracy, energy intensity and the overall control of the resulting focal region.

Here we show that a device based on highly nonlinear wave propagation can more accurately focus acoustic signals into a host medium, achieving a tunable, high amplitude focal region whose position is easily controllable. We design an acoustic lens composed of an array of nonlinear transducers based on discrete power-law materials (e.g., chains of spherical particles) (Fig. 1A). In contrast to linear elastic materials in which force $F$ and deformation $\delta$ obey the relation $F = k\delta$, discrete power-law materials do not support tensile forces and feature an unusual behavior described by $F = k\delta^n$, where $k$ is a stiffness constant and $n > 1$ a power-law exponent. The nonlinearity of the force-deformation relation makes chains of particles a host for unique acoustic waves that are stable, and whose behavior can easily be controlled (*9–13*). We take advantage of these characteristics to achieve enhanced control over sound focusing, de-focusing, and redirection. This approach offers a simple way to modify incoming acoustic waves into high-amplitude confined pulses (see insets in Fig. 1A).

Power-law materials like arrays of spherical particles can support linear, weakly nonlinear, and highly nonlinear wave dynamics, depending on the initial strain state of the material, which can be controlled by a static compression force $F_0$ (*9*) (Fig. 1B). For incoming signals leading to low particle interaction forces $F_m$ such that $F_m \ll F_0$, the resulting wave field is that of linear elastic materials. If $F_m \approx F_0$, the ensuing response is characterized by the classic, weakly nonlinear, soliton solution of the Korteweg-De Vries (KDV) equation (*9, 14*), in which the wavelength depends on the phase velocity $V_s$ (*15*). The weakly non-linear regime is depicted as the gray shaded area in Fig. 1B.

If $F_m \gg F_0$, the wave field is highly nonlinear and supports compact solitary waves or combinations thereof, in which the wavelength is constant (approximately five sphere diameters) and the phase velocity is dependent on amplitude (*9, 10*). In the continuum limit (wavelength is long compared to the sphere diameter), the highly nonlinear solitary waves have the compact structure

$$u(x,t) = \begin{cases} \cos^4\left[k_s\left(x - V_s t\right)\right] & k_s\left(x - V_s t\right) \in [-\pi/2, \pi/2] \\ 0 & otherwise \end{cases} \tag{1}$$

Here $u$ is the displacement of a material point, and $x, t$ are the spatial and temporal coordinates. For a given particle size, the wavenumber $k_s$ is a constant. Input signals of short duration lead to a solitary wave, while inputs of longer duration lead to a train of solitary waves or shock-like structures (see insets in Fig. 1A). All three stationary waveforms travel with a phase velocity $V_s$ that depends on the maximum dynamic force $F_m$ and the static compressive force $F_0$ (*13,15,16*). The frequency content of emitted waves is proportional to $kV_s$ and thus is easily controlled by properly selecting $F_0$. This can be used to control the frequency content of the acoustic field at the focal point.

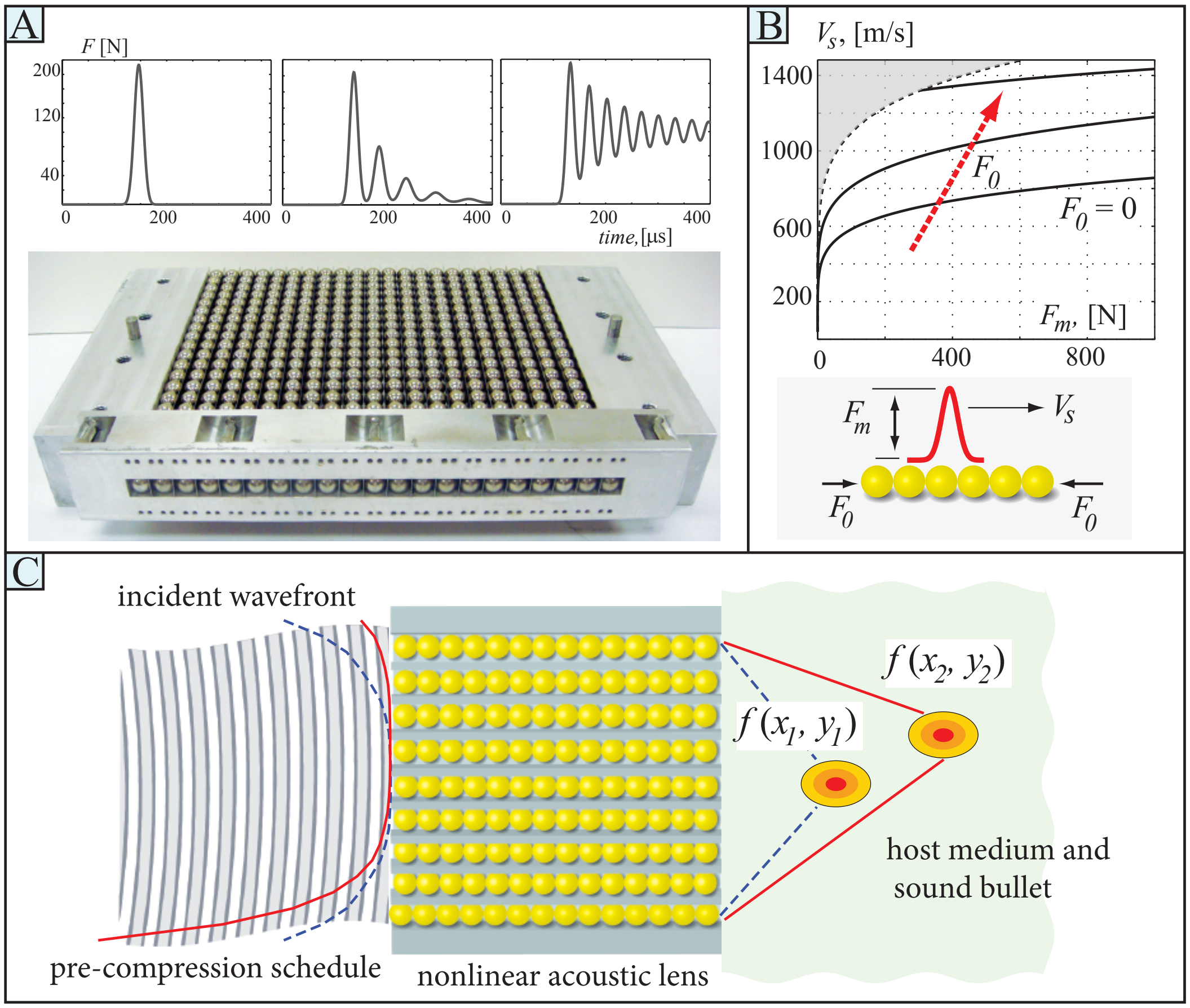

Figure 1: Prototype of a nonlinear acoustic lens **(A)** with 21 independent chains of spheres, each composed of 21 steel spheres and employed as the wave-modulation component (density $\rho = 8100$ Kg/m$^3$, Young's modulus $E = 196$ GPa, and diameter $D = 9.5$ mm), contained in an aluminum casing. The top insets depict a numerically simulated single solitary wave (left), a train of solitary waves (center), and a shock (right) traveling in a chain of steel spheres. The phase velocity $V_s$ strongly depends on the initial strain state determined by static compression $F_0$ **(B)** (the shaded gray region indicates a weakly nonlinear response). These characteristics lend themselves for the implementation of devices capable of focusing, de-focusing, or redirecting incoming waves **(C)**. A parabolic distribution of $F_0$ may be used to focus acoustic energy into a "sound bullet". The position of the focus lies on the lens' symmetry axis $(f(x_1, y_1))$ when $F_0$ is symmetrically distributed according to the dashed blue line, or otherwise off the symmtry axis $(f(x_2, y_2))$ for an asymmetric distribution as for the solid red line.

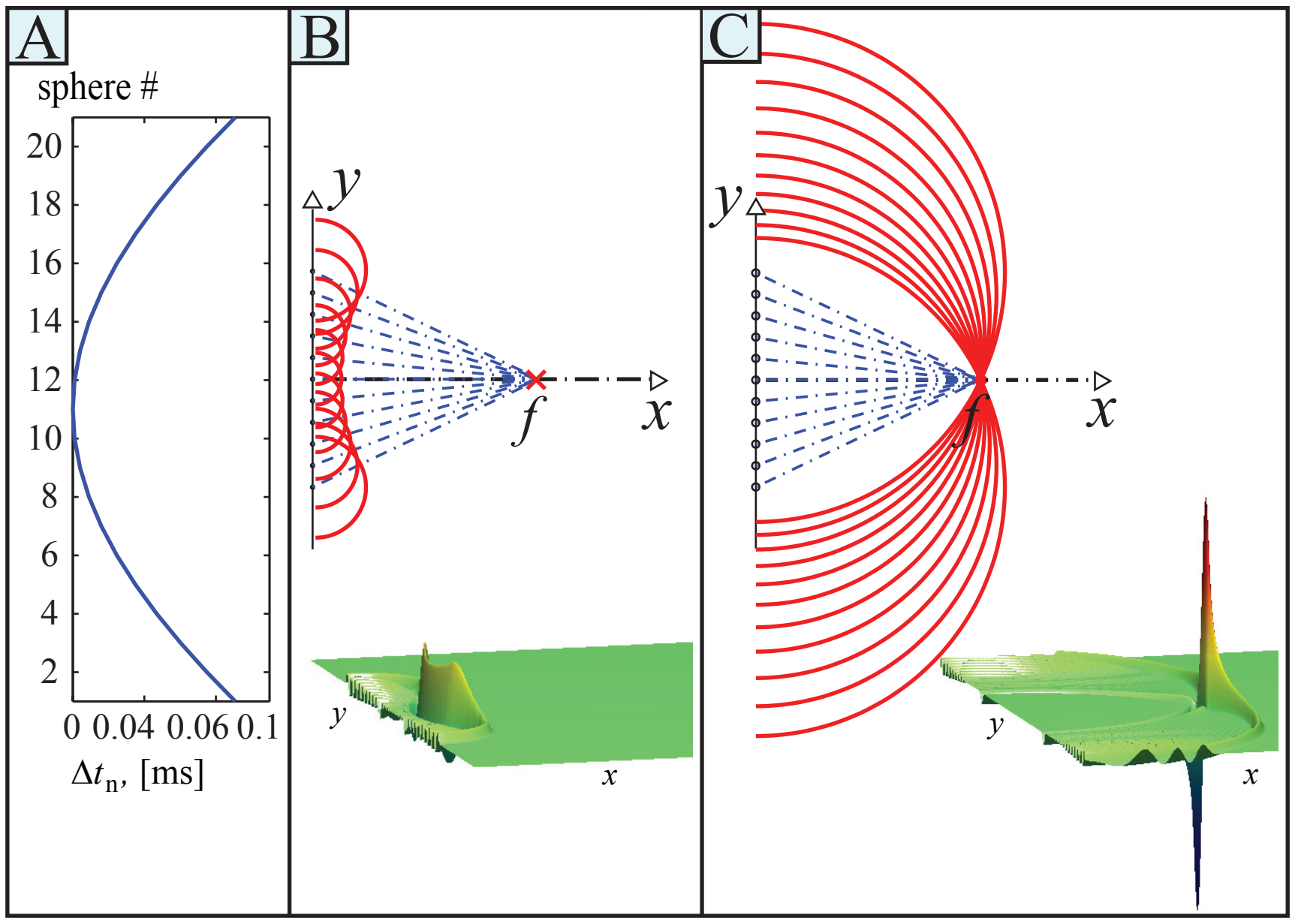

Figure 2: Idealized representation of spherical waves emitted by sources distributed on a plane, with a parabolic time delay distribution **(A)** necessary to obtain a focal point along the lens' symmetry axis; waves and pressure field from Eq. (2) shortly after generation **(B)**, and coalescing at the focal point **(C)**. Dashed blue lines indicate the distance of each source from the focal point while red solid lines denote the wave crest of each spherical wave. The resulting sound bullet is composed of symmetric maxima and minima, as shown in the 3D wave field plot on each panel.

In our theoretical analysis, we consider the focusing properties of the nonlinear acoustic lens in solid and fluid host media, limiting the study to the case of $F_m \gg F_0$. The interactions between the individual granular chains and the host medium is given by the contact of solid spheres with either a solid half space or thin metal plates separating the spheres themselves from a fluid half space.

In both cases, we assume the mechanical disturbances to be perceived as point sources by the host linear isotropic medium. Geometric or ray acoustics (*15*) can then be used to estimate the delay $\Delta t$ distribution necessary to focus energy at a desired location $(x_f, y_f)$ (*17*).

In the case of a fluid host medium, assuming that the displacement of each thin metal plate is represented by the same function as that describing a solitary wave (Eq. (1)) with arbitrary amplitude $A_n$ yields the pressure field (*17*)

$$p(x,y,t) = \begin{cases} \frac{\rho c k_s}{2} \sum_{n=1}^{N} \frac{A_n[2\sin(-2k_s\phi_n)+\sin(-4k_s\phi_n)]}{r_n} & k_s\phi_n \in [-\pi/2, \pi/2] \\ 0 & otherwise \end{cases} \quad (2)$$

where $N$ is the number of chains composing the lens, $r_n = \sqrt{(x-x_n)^2 + (y-y_n)^2}$, $k_s = V_s\sqrt{10}/(5D)$, and $\phi_n = t - r_n/c - \Delta t_n$ (*17*).

We designed a lens and its pre-compression schedule (*17*) to obtain a focal point at ($x_f = 13$ cm, $y_f = 0$ cm) in air (see Tables S1 and S2): the required time delay (from Eq. (S1)) is shown in Fig. 2A. The emitted spherical waves (Fig. 2B) coalesce at the focal point to produce a compact and high-amplitude pressure locus–a sound bullet (Fig. 2C). At the time of wave coalescence $t_0$, the focal region features a symmetric pressure distribution with one maximum and one minimum. The geometric superposition of incoming compact waves also allows the high pressure region to travel for controllable distances at the speed of sound in the host medium. The benefits offered by the proposed acoustic lens become apparent when compared to an array of linear sources driving the baffle as $\cos(\omega t)$, where $\omega = k_s V_s$. For the same conditions above, our nonlinear lens provides a focused signal in the host domain that is 6.4 times more compact than the same obtained with linear sources (*17*).

To test the validity of the assumptions regarding the interface of the acoustic lens with air and the evolution of the spherical waves in the host medium, we coupled a fluid-structure-interaction (FSI), finite-element (FE) model to a discrete-particle (DP) model (Fig. S1). The FSI model is employed to evaluate the interaction of moving thin metal plates and the host fluid domain. For simplicity, the pressure at the boundaries is assumed zero (corresponding to assuming the fluid to be confined to a closed box), explaining the reverberation observed in Fig. 3A, and both air and the particles in the lens are assumed initially at rest. The DP model, moreover, considers each sphere as a point mass, and the contact between any two spheres is treated as a spring of stiffness determined by Hertz contact theory (*9–11, 13, 18*). The simulated acoustic lens is identical to the prototype illustrated in Fig. 1A and composed of 21 chains of 21 spheres. We excite acoustic waves by impact. For a striker with initial velocity of 1 m/s and mass equal to that of 21 spheres, the ensuing pressure field in air for a desired focal location ($x_f = 9$ cm, $y_f = 7$ cm) is shown in Fig. 3A along with the required time delay. The sound bullet, formed off axis because of the selected asymmetric pre-compression schedule, attains a maximum pressure $p_B \approx 79$ Pa corresponding to $38$ dB. The pressure distribution at the focal point, shown as an inset in Fig. 3A, qualitatively agrees with that predicted by Eq. 2 obtained from first-principle considerations. The distance traveled by the sound bullet is evident in the root-mean-square (RMS) pressure field (Fig. 3B). Remarkably, the sound bullet travels for a finite distance, several times the signal wavelength, while maintaining its compact structure.

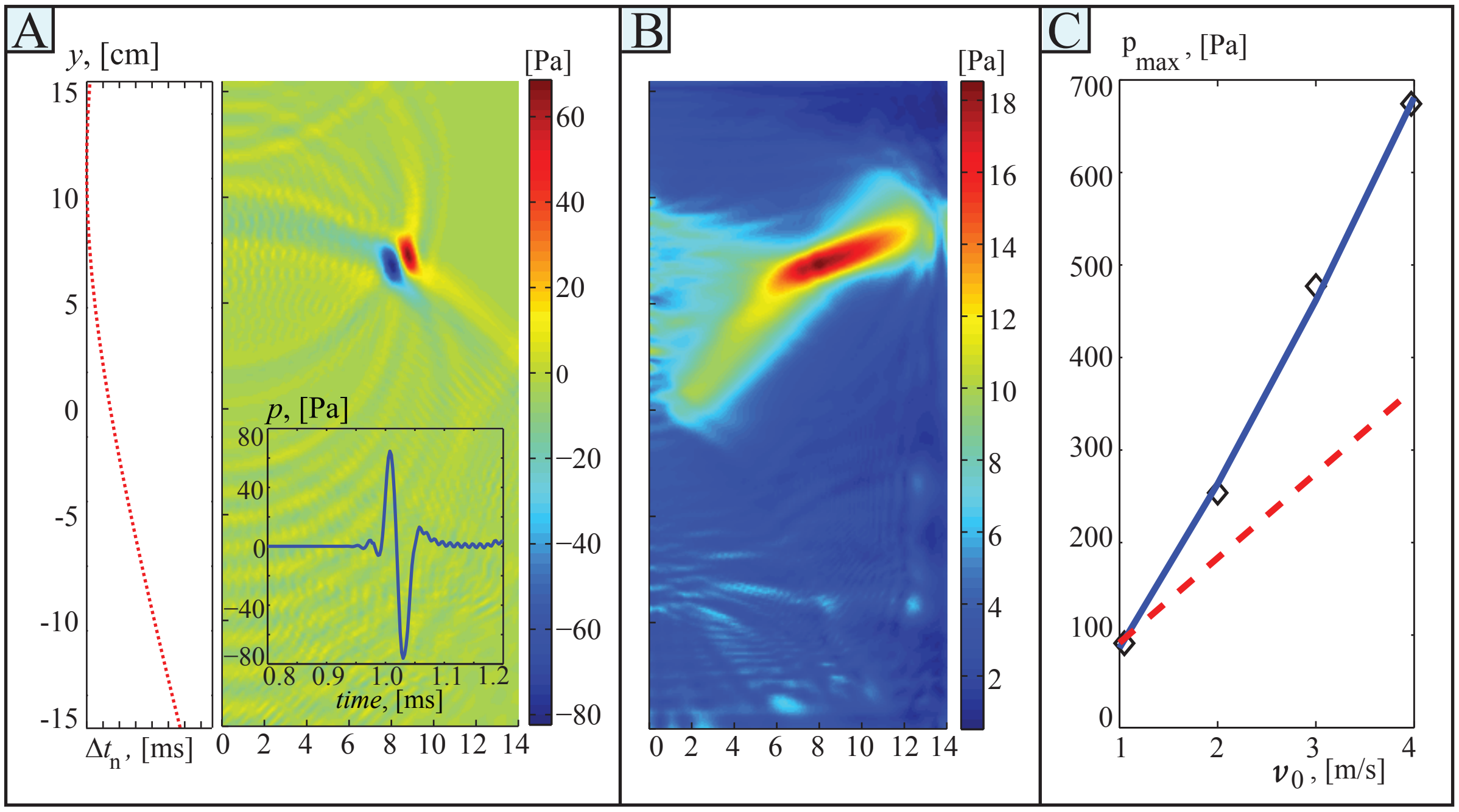

Figure 3: Interaction of the nonlinear acoustic lens with air computed via a fluid-structure-interaction, finite-element model. The sound bullet in air away from the lens symmetric axis **(A)** results from an asymmetric delay distribution. The pressure history at focal point ($x_f = 9$ cm, $y_f = 7$ cm) shown in the inset produces the same distribution qualitatively predicted with Eq. (2). The root-mean-square (RMS) pressure field **(B)** illustrates the distance traveled by the sound bullet, which remains compact. Increasing the impact velocity $v_0$ leads to a nonlinear sound-bullet pressure increase **(C)**, where the dashed red line indicates a linear relation between $v_0$ and focal pressure, the black diamonds indicate the computed values fitted by the solid blue line.

The balance between nonlinearity and dispersion in the chains of spheres, moreover, results in the generation of stable and compact solitary waveforms (*9,10*). This characteristic is retained even if the input amplitude is increased, leading to arbitrarily large signals within the sphere chains provided the force-deformation relation is not altered. The pressure amplitude of sound bullets increases super linearly as the impact velocity $v_0$ increases. If $v_0$ is set to 4 m/s, the resulting sound bullet attains a maximum pressure of 675 Pa corresponding to 57 dB. The relation between impact velocity and resulting pressure at the focal point is nonlinear (Fig. 3C).

We performed experiments to evaluate the focusing of pressure waves within a solid medium, and extended the analytical model and combined DP/FE simulations to support and validate the experimental data (*17*). We constructed a nonlinear acoustic lens prototype (Fig. 1A) and rested it on a polycarbonate plate as the host medium. The static precompression on the lens was regulated via variable weights attached to a fishing line threaded through the first sphere of each chain. An impacting mass was dropped on the top of the lens to excite compressional waves in the sphere chains. The resulting waves propagating in the sphere chains had a shock-like profile as computed with the DP/FEM model (Fig. 4B).

The waves emitted by the acoustic lens were delayed by the chosen pre-compression to obtain a symmetric focal point $10$ cm from the upper edge of the plate(Fig. 4A). The resulting dynamic stress transmitted in the plate produced photoelastic fringes that were captured with a black-and-white high speed camera and compared to the corresponding DP/FE model. The image acquisition was compared with the computed stress in the $y$-direction $\sigma_y$ at the focal location (Fig. 4C). The results show good qualitative agreement (Fig. 4D). Subtle misalignment of the striker, spheres and mid-plane of the plate produced other wave modes that polluted the photoelastic fringes experimentally measured. In spite of this, a clear stress concentration is observed, reaching the focal point with only a $\approx 5\%$ time delay between experiments and the numerical prediction (Fig. 4D).

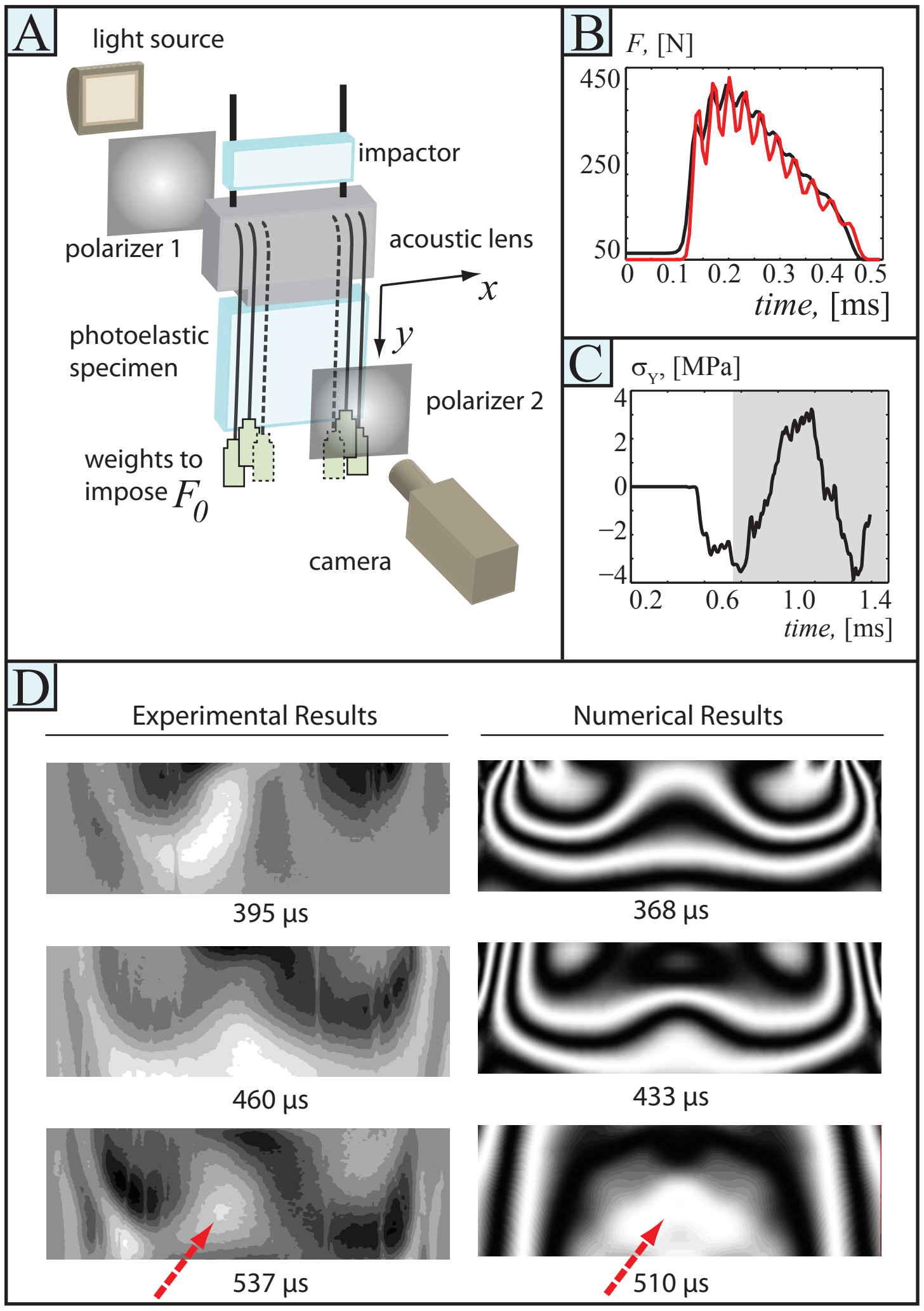

Figure 4: Schematic representing the photoelastic experiment employed to measure stress focusing within a polycarbonate plate **(A)**. The waves within the sphere chains are computed to be shock-like structures **(B)** due to the large impacting mass used the input. The red line corresponds to waves in the most compressed chain (1 and 21). The black line indicates waves in the uncompressed chain (11). The stress component $\sigma_y$ at the focal point **(C)** shows the coalescing of pressure waves at $t = 0.58$ ms. The area shaded in gray indicates the presence of reflected waves from the bottom portion of the plate. Photoelastic fringes and equivalent field computed with a FE model **(D)** provide qualitative agreement. The arrows point at the location of the focal point.

The studied highly nonlinear lens system expands the capabilities of existing acoustic transducers. It transforms a given acoustic excitation into either a single pulse or trains of pulses (compact solitary waves) that are highly tunable via mechanical means. The phase velocity of the waves is directly controlled by adjusting the static compression on each chain of spheres composing the lens, while the wavelength is determined by the size of the particles alone. These characteristics allow the focusing of waves transmitted into a host medium to generate compact sound bullets of arbitrarily large amplitudes. These sound bullets can travel in the host medium for relatively long distances while retaining a compact shape. This is currently not achievable by available technologies, which only allow oscillatory signals. Given the stability and compact nature of the highly nonlinear waves traveling in the granular chains, the proposed acoustic lens can be used as an effective mechanical filter, capable of controlling the frequency content of the resulting pressure field at the focal point. It features a very broad operational envelope, as linear, weakly nonlinear, and highly nonlinear behavior can be attained. For the considered conditions the proposed lens allows to obtain a focal area that is 6.4 times smaller compared to an ideal, linear device. The amplitude of the sound bullet can be significant, as the signals within the lens can be arbitrarily large. The described system has the potential to dramatically impact a variety of applications such as biomedical devices, non-destructive evaluation techniques, and defense systems.

19. The authors would like thank Dr. Veronica Eliasson and Mr. Tian Lan for their support and helpful discussions regarding experiments. Support from NSF MRSEC at Caltech and NSF CAREER (CD) is also acknowledged.

Supporting Online Material

www.sciencemag.org

Materials and Methods

Tables S1 and S2

Figs. S1, S2, S3

Eqs. (*S1*-*S12*)

# Supporting Online Material for Generation of Sound Bullets with a Nonlinear Acoustic Lens

Alessandro Spadoni, and Chiara Daraio*

*To whom correspondence should be addressed; E-mail: daraio@caltech.edu.

## Materials and Methods

### Material Properties of Spheres and Neighboring Media

In all considered cases, the adopted spheres are made of stainless steel with material and geometric properties summarized in Table S1. The considered host media for acoustic focusing are air and polycarbonate (*S2*). The material properties for each are reported in Table S2.

Table S1: Geometric and material properties (stainless steel 302) from (*S1*) of employed spheres

| | |
|---|---|
| Diameter | $D = 9.5$ mm |
| Young's modulus | $E = 195.6$ GPa |
| Density | $\rho = 8100$ Kg/m$^3$ |
| Poisson's ratio | $\nu = 0.33$ |

### Time Delay via Static Compression

We assume the mechanical disturbances generated by the nonlinear lens to be perceived as point sources by the host medium, owing to the small contact area of a sphere and an adjacent planar surface. Such point sources, moreover, produce spherical acoustic waves that propagate unobstructed in the host medium, taken to be linear and isotropic. Geometric or ray acoustics (*S3*) is thus in order to estimate the delay $\Delta t$ distribution necessary to focus energy at a desired location $(x_f, y_f)$, which is:

$$c^2 \left(t_0 - \Delta t_n\right)^2 = \left(x_f - x_n\right)^2 + \left(y_f - y_n\right)^2 = r_n^2 \tag{S1}$$

where $(x_n, y_n)$ is the location of each $n$ source, $c$ is the speed of sound of the linear medium and $t_0$ is the time of flight between the farthest source and the focal point.

The time delay $\Delta t_n$ established above can be converted to a particular pre-compression distribution since the phase velocity $V_s$ is highly dependent upon the static compressive force $F_0$. Defining the quantity $F_r = F_m/F_0$ and knowing the length of a sphere chain, one can convert a required time delay $\Delta t_n$ with (*S4,S5*):

$$V_s = \frac{c_0}{(F_r - 1)} \left\{ \frac{4}{15} \left[ 3 + 2F_r^{5/3} - 5F_r^{2/3} \right] \right\}^{1/2}, \tag{S2}$$

where the characteristic "speed of sound" $c_0$ is (*S4*)

$$c_0 = \sqrt{\frac{E}{\rho}} \left[ \sqrt{\frac{81F_0}{\pi E}} \frac{1}{\pi D (1 - \nu^2)} \right]^{1/3}, \tag{S3}$$

Table S2: Material properties of air and Polycarbonate (*S2*)

| | |
|---|---|
| Air | |
| Density | $\rho = 1.225$ Kg/m$^3$ |
| speed of sound | $c = 343$ m/s |
| Polycarbonate | |
| Young's modulus | $E = 3.45$ GPa |
| Density | $\rho = 1230$ Kg/m$^3$ |
| Poisson's ratio | $\nu = 0.35$ |
| speed of sound | $c = 1675$ m/s |

In Eqs. S2 and S3, $\nu$ denotes the Poisson's ratio of the material, $E$ is the Young's modulus, $D$ is the spheres diameter, and $\rho$ is the material density. Note that $c_0$ is always lower than that of the material ($\sqrt{E/\rho}$) unless $F_0 \sim D^2 E$.

**Pressure field from first principles**

The solitary waves traveling along each chain of spheres in the acoustic lens are expected to generate displacements in each plate separating the nonlinear phased array and neighboring linear fluid of the following form:

$$u_w(t) = \cos^4 \left[ -\frac{\sqrt{10}}{5D} V_s (t - \Delta t) \right]. \tag{S4}$$

In eq. (S4), $u_w(t)$ denotes the normal component of displacement. The pressure field in the linear medium is now obtained by superposing the contribution of each plate, as if it were an independent source, in the form of boundary forcing. With the assumed form of the baffle motion, the near-field boundary condition at each interface between spheres, baffle and fluid is obtained from Eq. (S4) as

$$\left.\frac{\partial \hat{p}_n}{\partial r}\right|_{r\to 0} = -\rho\frac{\partial^2 u_{w,n}}{\partial t^2} = \rho A_n g_n(t-\Delta t_n), \tag{S5}$$

where $A_n$ is a constant related to the pressure amplitude at the boundary and $\hat{p} = p/r$. The function $g_n$ is $g_n = \partial^2 u_{w,n}/\partial t^2$. The absence of waves converging from infinity and the initial conditions

$$\hat{p}_n(r,0) = \dot{\hat{p}}_n(r,0) = 0 \tag{S6}$$

lead to a pressure field in Fourier domain of the form:

$$\hat{p}_n(x,y,t) = \frac{1}{\sqrt{2\pi}}\int_{-\infty}^{\infty}\left[\rho c A_n \frac{G_n(\omega)}{i\omega}e^{-i\omega\Delta t_n}e^{-i\omega r_n/c}\right]e^{i\omega t}d\omega, \tag{S7}$$

or

$$\hat{p}_n(x,y,t) = \int_0^{\phi_n}\rho c A_n g_n(\tau)d\tau, \tag{S8}$$

where $\phi_n = t - r_n/c - \Delta t_n$. Given the assumption of linearity, the pressure field within the fluid medium of interest is then

$$\hat{p}(x,y,t) = \frac{\rho c k_s}{2}\sum_{n=1}^{N}A_n\left[2\sin\left(-2k_s\phi_n\right) + \sin\left(-4k_s\phi_n\right)\right] \tag{S9}$$

where $k_s = V_s\sqrt{10}/(5D)$. Note that, for a single solitary wave emanating from each chain of spheres in the nonlinear acoustic lens, Eq. (S9) is defined in the same domain or $k_s\phi_n \in [-\pi/2, \pi/2]$. The pressure field outside this domain is zero, implying the compact support for the sound bullets. Eq. (*S9*) also contains information on the frequency content at the foci. Such frequency can be tuned varying the static pre-compression on the chains.

The pressure field for an array of $N$ linear sources driving each plate, separating the array and neighboring fluid, can be computed in an analogous manner. Assuming a baffle motion of the form

$$u_w(t) = \cos\left[\omega\left(t-\Delta t\right)\right] \tag{S10}$$

where $\omega = k_s V_s$, Eq. (S8) along with initial and boundary conditions indicated by Eqs. (S6) and (S5) yields the pressure field

$$\hat{p}(x,y,t) = \rho c k_s\sum_{n=1}^{N}A_n\sin\left(-k_s\phi_n\right) \tag{S11}$$

where it is assumed that $k_s\phi_n \in [-\pi,\pi]$ or a single period of the assumed waveform. This is optimistic since electromechanical transducers often lead to multiple-period signals, but it is nonetheless considered for the sake of comparison with our nonlinear lens.

For an impact of a striker with mass equal to that of 21 steel spheres with properties listed in Table S2 and an initial velocity $v_0 = 1$ m/s, the phase velocity $V_s$ of the solitary waves within the lens is approximately $658$ m/s and $k_s = 66.4$ rad/m.

### Combined Discrete-Particle (DP) and Finite-Element (FE) Model of the Acoustic Lens

The numerical analysis serves as tool to quantify the pressure field produced by the transmission of waves within arrays of spheres into the host medium. The fluid or solid space adjacent to the acoustic lens is considered linear, isotropic and modeled as 2D. A fluid-structure-interaction (FSI) model is established to analyze the interaction of a flexible baffle with the neighboring fluid. The baffle is discretized via Timoshenko beam elements, while coupling fluid-structure and fluid-only elements are plane, 4-node, isoparametric elements (*S6*). The acoustic lens is instead considered to be directly in contact with a solid half space, which is discretized via plane, 4-node, isoparametric elements as well. The choice if isoparametric elements is motivated by the fact that significant resolution is needed at the boundary between spheres and the host medium, while a coarser discretization may be reasonable in other areas. This scenario is most easily treated with an unstructured or free mesh. If the neighboring medium is a fluid (Fig. S1A), arrays of spheres interact with a baffle (end plates shown as orange lines in Fig. S1A), which in turn interacts with the adjacent fluid. The FSI model (*S6*) allows to simulate the coupled dynamics of the baffle and fluid. The interaction of the spheres with the baffle itself is modeled by imposing equilibrium of forces as shown in Fig. S1B. Spheres are considered point masses and their interaction is modeled by nonlinear springs (*S7*) whose stiffness constant is provided by Hertz contact theory (*S8*). If the neighboring medium is a solid, there is no interface baffle and the spheres contact the solid directly.

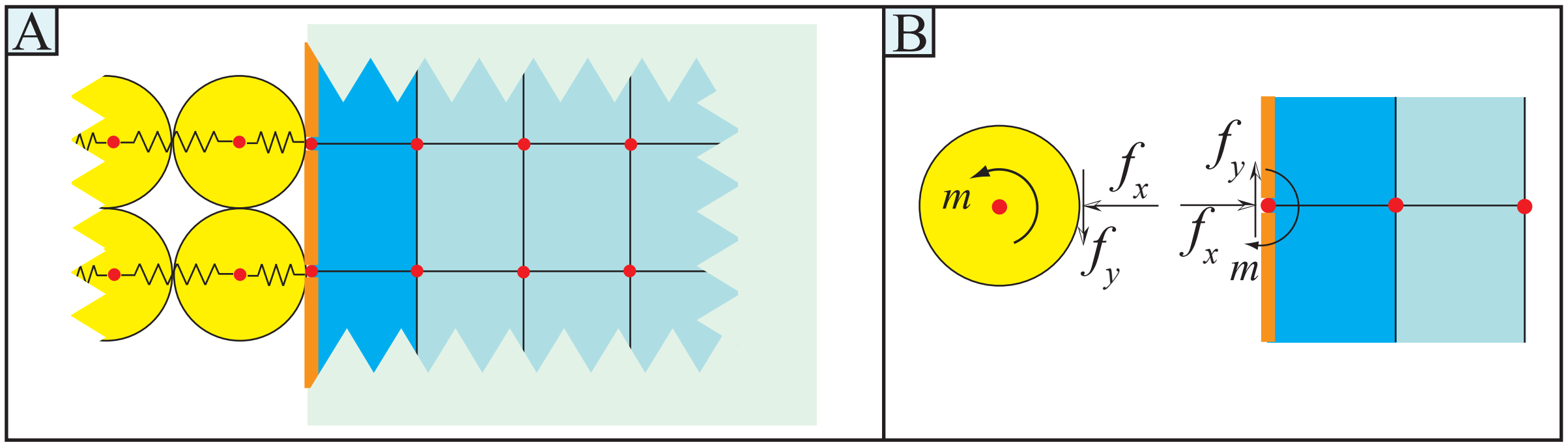

Figure S1: Combined DP/FE model. Arrays of spheres **(A)** interact with a baffle (orange lines) which in turn interacts with the adjacent fluid. The interaction of granular media and baffle **(B)**.

The mesh employed for the fluid case in shown in Fig. S2. In the fluid case, beam elements simulating the baffle are superposed to the plane elements. In the solid case, the same discretization is employed, but the solid domain has dimensions of the polycarbonate plate reported below.

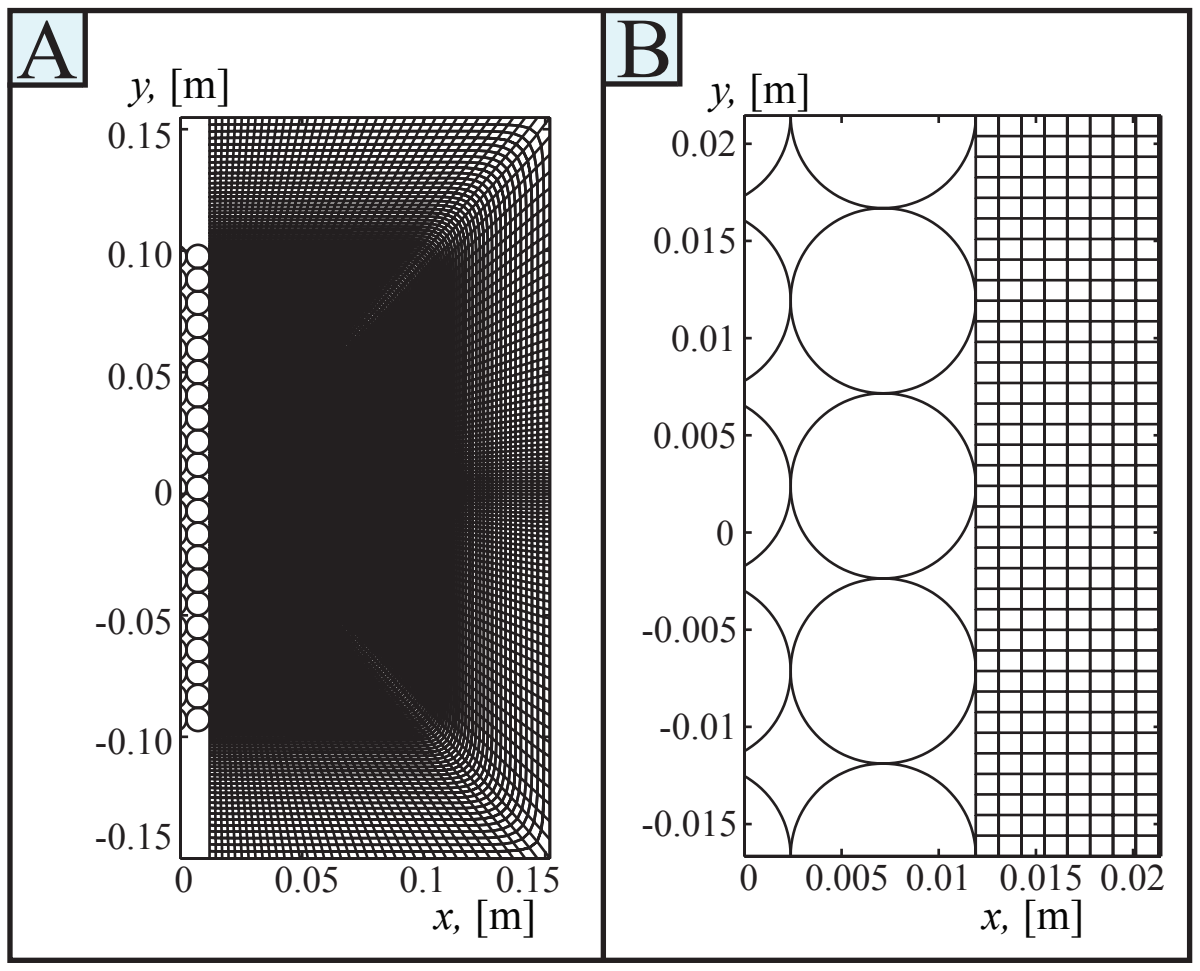

Figure S2: Discretization of host medium with spheres **(A)**, and detail of their interaction **(B)**.

To compare numerical and experimental models, the numerical stress field was converted to visible photoelastic fringes by the stress optic law (*S2*):

$$n = \frac{h}{f_\sigma} (\sigma_1 - \sigma_2) \tag{S12}$$

where $h$ is the polycarbonate-plate thickness and $f_\sigma = 7$ KN/m is the material fringe constant. The number of visible fringes $n$ is proportional to the difference of principal stresses $\sigma_1$ and $\sigma_2$.

**Experimental setup and Methods**

The experimental prototype of the acoustic lens (shown in Fig.1A) is made of Aluminum 6061-T051 and measures $30.5 \times 23 \times 7.6$ cm. It accommodates 21 chains each composed of 21 stain steel spheres (with properties listed in Table S1). Since each chain was statically compressed by a different force $F_0$, arrays of spheres were separated from neighboring chains with stainless steel shim stock of thickness $0.15$ mm. The inside surface of the lens cases were lined with Teflon sheets to minimize friction. The front and back sides of the lens casing feature slots (Fig. S3A) to accommodate individual strands of fishing line used to compress each chain. The top-most sphere of each chain was threaded with fishing line, and was secured to one side of the lens casing, to which weights were connected. Water bottles with variable water content were used as they provided significant resolution in setting the proper pre-compressive weight. The assembled lens was vertically rested on a polycarbonate plate (Table S2) of dimensions $25 \times 25 \times 1.9$ cm, which in turn rested on a table (Fig. S3A).

Upon loading of the sphere arrays, each chain assumed a distinct length (owing the distinct compressive force applied) posing a challenge in imparting mechanical energy to each chain simultaneously. Accordingly, this striker was made to coincide with the spheres protruding from the top portion of the lens. The adopted solution consisted in drilling holes in the first sphere of each chain to accommodate a 2.5-cm long screw. The impacting plate was modified by adding 21 holes (2.5 cm in depth, 1 cm in diameter) to accommodate the screws protruding from each sphere. The holes were then filled with a two-part CASTAMOUNT acrylic resin from Pace technologies. In this manner, the impacting plate feature a contact edge enhanced with protruding spheres of identical arrangement as the top portion of each chain (Fig. S3A). Finally, the camera was triggered using a PCB accelerometer placed on the top portion of the impactor assembly.

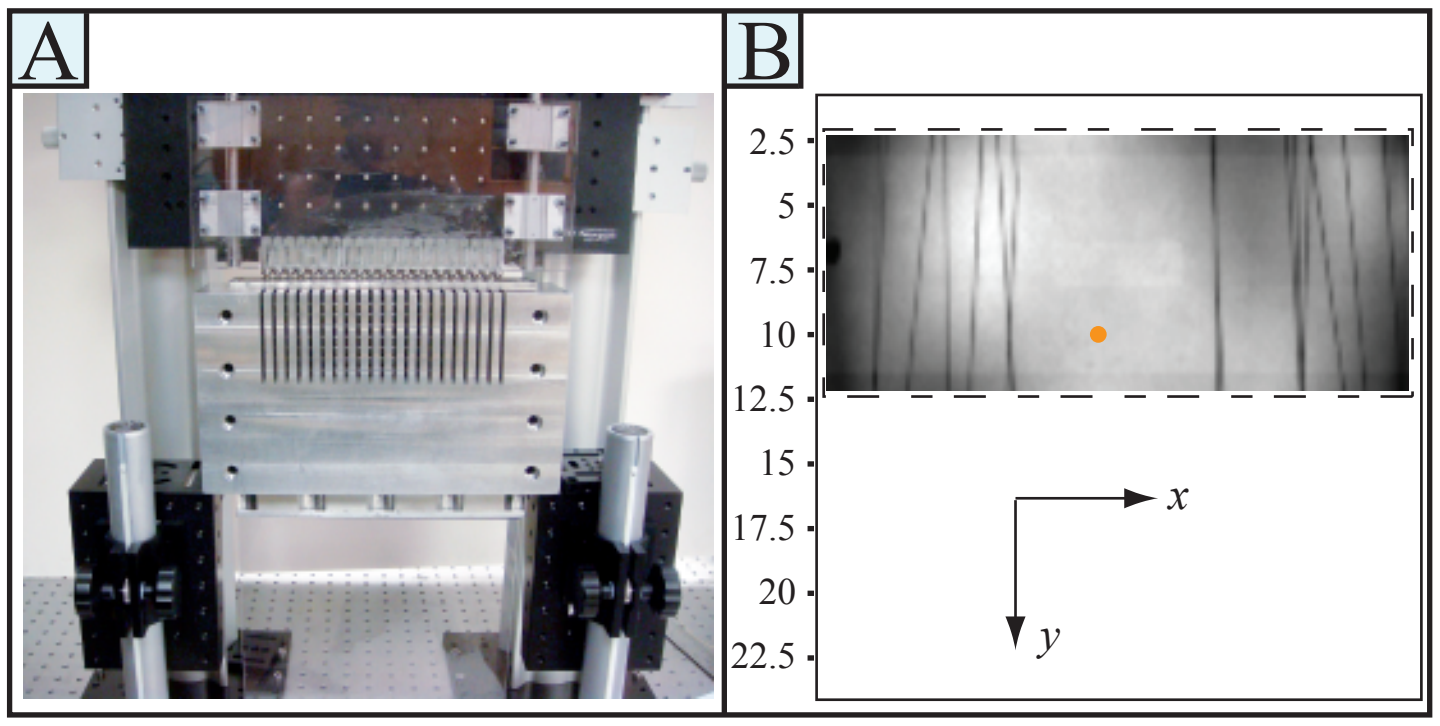

Figure S3: Experimental setup **(A)** with striker and acoustic lens vertically resting on a Polycarbonate plate of properties listed in Table S2. Holes filled with resin to properly align the striker edge with the top portion of the lens are visible. The striker is mounted onto two rails to ensure proper impact. The high-speed camera acquired images of a portion of the Polycarbonate plate **(B)**. The nominal focal point is indicated by the orange dot. The dark lines are shadows of the fishing line used to impart static compression to each chain.

Using the DP/FE model for guidance, a Vision Research, Phantom V12 high speed camera was triggered $280\ \mu$s after impact. The black-and-white camera was set to acquire images at $300,000$ samples per second. The same camera, furthermore, was made to acquire the portion of the Polycarbonate plate shown in Fig. S3B, which measured approximately $25 \times 10$ cm. In order to remove the shadows from the fishing line and enhance the visible photoelastic fringes, the reference image (Fig. S3B) was subtracted from each acquired frame.